\documentclass[prb,twocolumn,floatfix,superscriptaddress,showpacs]{revtex4}
\usepackage{graphicx,color}
\usepackage{amsmath,amssymb,bm,bbm}

\newcommand{\mc}{\mathcal}

\newcommand{\tn}{\textnormal}
\newcommand{\cprb}[3]{Phys.~Rev.~B {\bf #1}, #2 (#3)}
\newcommand{\cprl}[3]{Phys.~Rev.~Lett.~{\bf #1}, #2 (#3)}

\newcommand{\cjp}[3]{J.~Phys.: Condensed Matter {\bf #1}, #2 (#3)}

\newcommand{\cbook}[2]{\textit{#1} (#2)}

\definecolor{darkred}{rgb}{0.90,0,0}
\definecolor{darkgreen}{rgb}{0,0.60,.2}
\definecolor{darkblue}{rgb}{0,0,1}
\definecolor{grey}{cmyk}{0,0,0,0.25}
\definecolor{orange}{cmyk}{0,0.6,0.8,0}

\begin{document}
\title{Finite-temperature linear conductance from the Matsubara Green function\\ without analytic continuation to the real axis}

\author{C.\ Karrasch}
\affiliation{Institut f\"ur Theorie der statistischen Physik and JARA -- Fundamentals of Future Information Technology, RWTH Aachen University, 52056 Aachen, Germany}
\author{V.\ Meden}
\affiliation{Institut f\"ur Theorie der statistischen Physik and JARA -- Fundamentals of Future Information Technology, RWTH Aachen University, 52056 Aachen, Germany}
\author{K.\ Sch\"onhammer}
\affiliation{Institut f\"ur Theoretische Physik, Universit\"at G\"ottingen, Friedrich-Hund-Platz 1, 37077 G\"ottingen, Germany}

\begin{abstract}
We illustrate how to calculate the finite-temperature linear-response conductance of quantum impurity models from the Matsubara Green function. A continued fraction expansion of the Fermi distribution is employed which was recently introduced by Ozaki [\cprb{75}{035123}{2007}] and converges much faster than the usual Matsubara representation. We give a simplified derivation of Ozaki's idea using concepts from condensed matter theory and present results for the rate of convergence. In case that the Green function of some model of interest is only known numerically, interpolating between Matsubara frequencies is much more stable than carrying out an analytic continuation to the real axis. We demonstrate this explicitly by considering an infinite tight-binding chain with a single site impurity as an exactly-solvable test system, showing that it is advantageous to calculate transport properties directly on the imaginary axis. The formalism is applied to the single impurity Anderson model, and the linear conductance at finite temperatures is calculated reliably at small to intermediate Coulomb interactions by virtue of the Matsubara functional renormalization group. Thus, this quantum many-body method combined with the continued fraction expansion of the Fermi function constitutes a promising tool to address more complex quantum dot geometries at finite temperatures.
\end{abstract}

\maketitle

\section{Introduction}
\label{sec:intro}
Computing the finite-temperature $T$ linear conductance for widely used model systems like an Anderson impurity connected to non-interacting leads involves integrals over the impurity single-particle spectral function $\rho(\omega)$ multiplied by the derivative of the Fermi distribution.\cite{pinkbook} At $T>0$, however, most quantum many-particle methods are set up in imaginary time and yield the relevant one-particle propagator $\mc G(i\omega_n)$ at the Matsubara frequencies $\omega_n$ on the imaginary axis. Thus, an analytic continuation is required in order to obtain the corresponding spectral functions on the real axis and to compute transport properties at finite temperatures. If $\mc G(i\omega_n)$ is only known numerically -- which is the generic scenario -- the analytic continuation can in principle be achieved, e.g., by Pad\'e approximation of the Matsubara data,\cite{pade1,pade2} but in general such a procedure is ill-controlled (see, e.g., Ref.~\onlinecite{frequenzen}). Alternatively, one can use the analytic properties of the Fermi function and resort to the calculus of residues to perform the integration directly on the imaginary axis. Even though this might involve the propagator $\mc G$ (or, aiming at the conductance, its derivative) not only at the physical Matsubara frequencies but also at different arguments, computing such data by \textit{interpolation} is expected to be significantly more stable than carrying out an analytic \textit{continuation}.

In order to rotate the path of integration from the real to the imaginary axis, one can employ the Matsubara expansion for the Fermi distribution $f(x)=1/[\exp(x) + 1]$, but this is not useful for practical applications in fermionic many-body problems because of its slow convergence.\cite{FW} A much more efficient representation was recently proposed by Ozaki.\cite{Ozaki} It is based on a continued fraction expansion and ultimately yields an infinite sum over simple poles which are all located on the imaginary axis. In its infinite form, this series is equivalent to the usual Matsubara expansion. Truncation, however, modifies the position of the poles (but they remain on the imaginary axis) as well as the corresponding residues, and it turns out that the resulting finite series yields an excellent approximation to $f(x)$ for not too large values of $|x|$, even if only a few terms are accounted for (all this will be quantified below).

The motivation for the present work is three-fold. First, we give a simplified derivation of Ozaki's idea which relies on concepts taken from condensed matter theory and present results for the rate of convergence of the truncated series (Section \ref{sec:fermi}). Secondly, we employ an exactly-solvable model -- a non-interacting tight-binding chain with a single site impurity -- and demonstrate that interpolation of the Matsubara Green function is much more stable than continuation to real frequencies, illustrating that it is advantageous to perform calculations on the imaginary axis (Section \ref{sec:pade}). Thirdly, the formalism is applied to the single impurity Anderson model, and the Coulomb interaction $U$ is tackled by an approximation scheme which (in thermal equilibrium) is most easily set up in Matsubara space -- the functional renormalization group (FRG).\cite{salmhofer,dotsystems,frequenzen,severinsiam} By comparison with very accurate numerical renormalization group (NRG) data, we demonstrate that the FRG allows for reliably extracting the finite-temperature linear conductance at small to intermediate values of $U$ (Section \ref{sec:siam}). This was previously not possible due to the instability of the analytic continuation.\cite{frequenzen} Whereas within the NRG framework the numerical effort grows exponentially when considering general models with several correlated degrees of freedom and one is generically restricted to situations of high symmetry, the Matsubara FRG can easily be generalized to multi-dot or multi-level quantum dot geometries without specific symmetries with only a mild increase in computation time.\cite{frequenzen} Combined with the continued fraction expansion of the Fermi function, the FRG thus holds the promise for future applications addressing transport properties of such systems at $T>0$.

\section{Continued-fraction representation of the Fermi function}
\label{sec:fermi}

\subsection{Derivation}
In this Section we give a simplified derivation of Ozaki's idea.\cite{Ozaki} The first step is to express the Fermi distribution in the form 
\begin{equation} \label{eq:fermi}
f(x)=\frac{1}{e^x+1}=\frac{1}{2}\left(
1-\frac{\sinh{(x/2)}}{\cosh{(x/2)}}\right)~,
\end{equation}
where later on $x=\beta(\omega-\mu)$, with $\beta$ and $\mu$ denoting the inverse temperature and the chemical potential, respectively. In order to obtain a continued fraction expansion for $\tanh(x/2)$, Ozaki introduced
the (generalized hypergeometric) auxiliary functions
\begin{equation}\label{eq:series}
F(a,x)= \sum_{n=0}^{\infty}\frac{x^n}{n!a_{(n)}}~,
\end{equation}
with $a_{(0)}=1$ and
\begin{equation}
a_{(n)}=a(a+1)\ldots(a+n-1)~,~~~~n\geq1~.
\end{equation}
It is straightforward to show that
\begin{equation} \label{eq:example}
\cosh{x}=F\left(\frac{1}{2},\frac{x^2}{4}\right)~,~~~~
\sinh{x}=xF\left(\frac{3}{2},\frac{x^2}{4}\right)~.
\end{equation}
The first identity, e.g., is easily confirmed by establishing $(2n)!/(4^nn!)=(1/2)_{(n)}$ by induction. Thus, $\tanh(x)$ is determined by a ratio of the form $F(a,x)/F(a-1,x)$, rendering it reasonable to seek for a recurrence relation. To this end, one calculates the difference of the power series of Eq.~(\ref{eq:series}) for $a-1$ and $a$:
\begin{eqnarray}
F(a-1,x)-F(a,x)
=\frac{x}{a(a-1)}F(a+1,x)~.
\end{eqnarray}
Dividing by $F(a,x) $, taking the inverse and replacing $a$ by $a_0+n$ gives
\begin{eqnarray}\label{eq:rr}
r_n(a_0,x)  =
\frac{1}{1+\frac{x}{(a_0+n)(a_0-1+n)} r_{n+1}(a_0,x)}~,
\end{eqnarray}
where we have introduced the ratio
\begin{equation}
r_n(a_0,x) = \frac{F(a_0+n,x)}{F(a_0-1+n,x)}~.
\end{equation}
If one starts with $n=1$, iteration yields an expansion for $r_1(a_0,x)$ in terms of a continued fraction. From Eqs.~(\ref{eq:fermi}) and (\ref{eq:example}), the corresponding representation for $[f(x)-1/2]/x$ is obtained setting $a_0=1/2$ and replacing $x$ by $x^2/16$.

\begin{figure}[t]
\centering
\includegraphics[width=0.9\linewidth,clip]{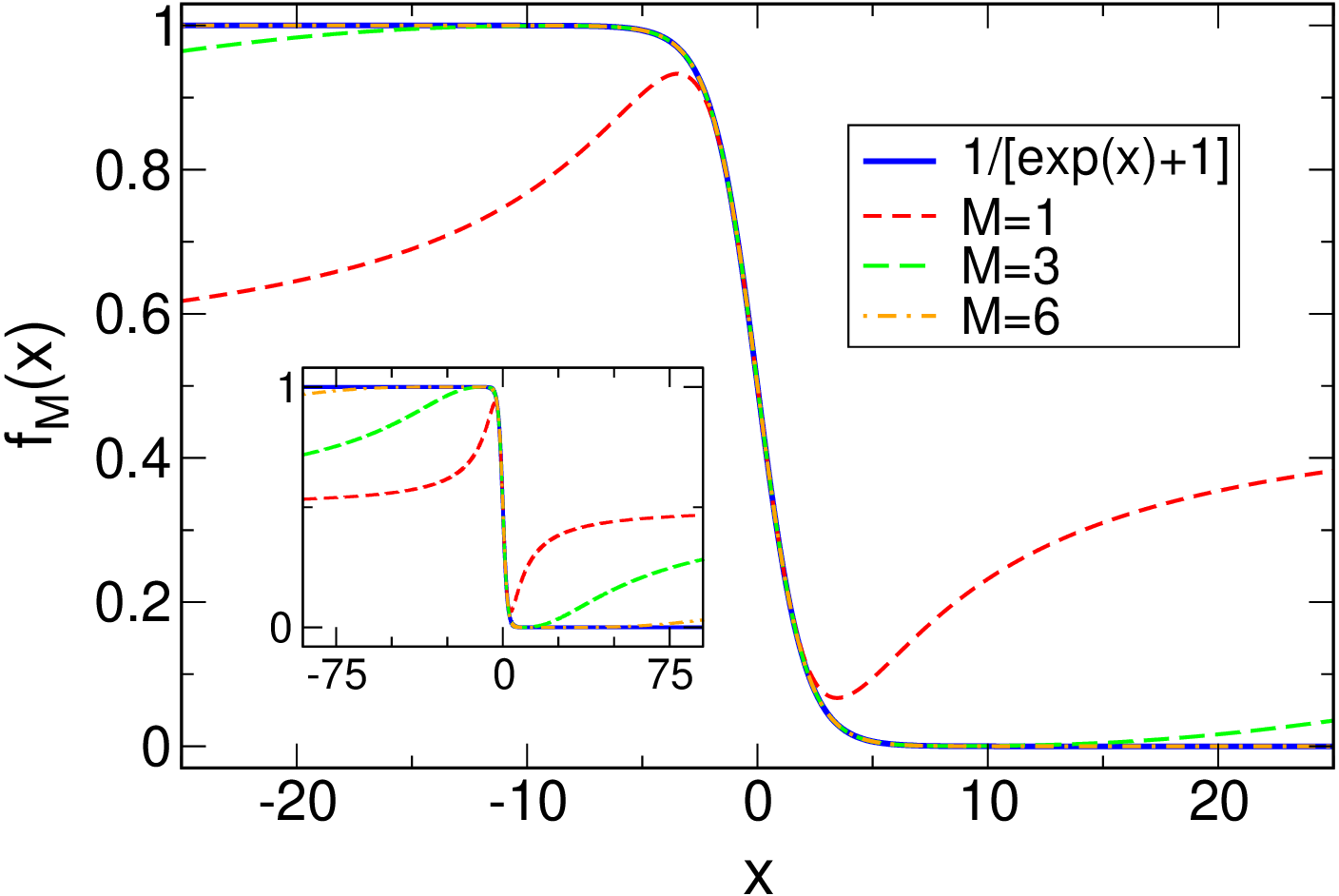}
\caption {(Color online) Approximation $f_M(x)$ to the Fermi function obtained from the continued fraction representation of Eq.~(\ref{eq:result2}) truncated at different $N=2M$. Already for $M=3$, the agreement with the exact result $f(x)=1/[\exp (x)+1]$ is excellent for not too large arguments $|x|\lesssim10$. Likewise, the $M=6$ - curve nicely approximates $f(x)$ up to $|x|\lesssim60$. For larger arguments, all $f_M(x)$ tend to $1/2$. }
\label{fig:fermi}
\end{figure}

For a practical application it is not necessary to write down the continued fraction explicitly. Rather, one can directly exploit the fact that the very same recursion relation of Eq.~(\ref{eq:rr}) appears if one aims at computing the inverse of an infinite tridiagonal matrix $C$ associated with a semi-infinite one-dimensional tight-binding chain with nearest neighbor hopping and starting at site $n=1$. Using the well-known Feshbach projection method,\cite{taylor} one obtains (for $n\geq 1$)
\begin{equation}
\left(C_{[n]}^{-1}\right)_{n,n}=\frac{1}{C_{n,n}-C_{n,n+1}\left(C_{[n+1]}^{-1}\right)_{n+1,n+1}C_{n+1,n}}~,
\end{equation}
where $C_{[n]} $ denotes the part of the matrix $C$ projected onto the sites greater and equal to $n$. Comparison with Eq.~(\ref{eq:rr}) and eventually combining with Eqs.~(\ref{eq:fermi}) and (\ref{eq:example}) yields
\begin{equation}\label{eq:result}
f(x)=\frac{1}{2}-\frac{x}{4}\langle 1|(1+ixB)^{-1}|1\rangle~,
\end{equation}
with the only non-vanishing matrix elements of $B$ (which decrease like $1/(4n)$ for large $n$) given by
\begin{equation}\label{eq:b}
B_{n,n+1}=B_{n+1,n}=\frac{1}{2\sqrt{(2n-1)(2n+1)}}~,~~~~n \geq1~.
\end{equation}
Thus, one finally faces the ordinary eigenvalue problem
\begin{equation}\label{eq:ew}
B|b_\alpha\rangle =b_\alpha |b_\alpha\rangle~,~~~~b_\alpha\in\mathbb{R}~.
\end{equation}
All $b_\alpha$ come in pairs owing to the fact that the state $|\bar b_\alpha\rangle$ with components $\langle n|\bar b_\alpha\rangle =(-1)^n\langle n|b_\alpha\rangle$ is an eigenvector of $B$ with eigenvalue $-b_\alpha$. Then, Eq.~(\ref{eq:result}) can be recast as
\begin{equation}\label{eq:result2}\begin{split}
f(x)-\frac{1}{2}&=-\frac{x}{4}\sum_\alpha\frac{|\langle  1|b_\alpha\rangle|^2}{1+ixb_\alpha}~\\
&=-\sum_{\alpha>0}\left [\frac{R_\alpha}{x-i/b_\alpha}  +\frac{R_\alpha}{x+i/b_\alpha}  \right]~,
\end{split}\end{equation}
with $R_\alpha= |\langle 1|b_\alpha\rangle|^2/(4b_\alpha^2)$, and $\alpha>0$ symbolizing that the sum extends over the positive part of the spectrum (i.e., all $\alpha$ for which $b_\alpha>0$) only. This expansion of the Fermi function should be compared to the well-known ($\beta=1$) Matsubara sum decomposition\cite{FW,Ozaki}
\begin{equation}\label{eq:resultmatsubara}
f(x)-\frac{1}{2}=
-\sum_{n=1}^\infty \left [\frac{1}{x-i\pi(2n-1)} 
 +\frac{1}{x+i\pi(2n-1) }  \right]
\end{equation}
which is slowly converging with w.r.t.~$n$. \textit{If the infinite matrix $B$ is not replaced by a finite one}, Eqs.~(\ref{eq:result2}) and (\ref{eq:resultmatsubara}) should coincide, implying that $R_\alpha=1$ and the positive eigenvalues given by $b_\alpha=1/[\pi(2n-1)]$. The fact that the latter accumulate at zero is obvious from the large $n$ decay of the matrix elements $B_{n,n+1}$.

\begin{figure}[t]
\centering
\includegraphics[width=0.9\linewidth,clip]{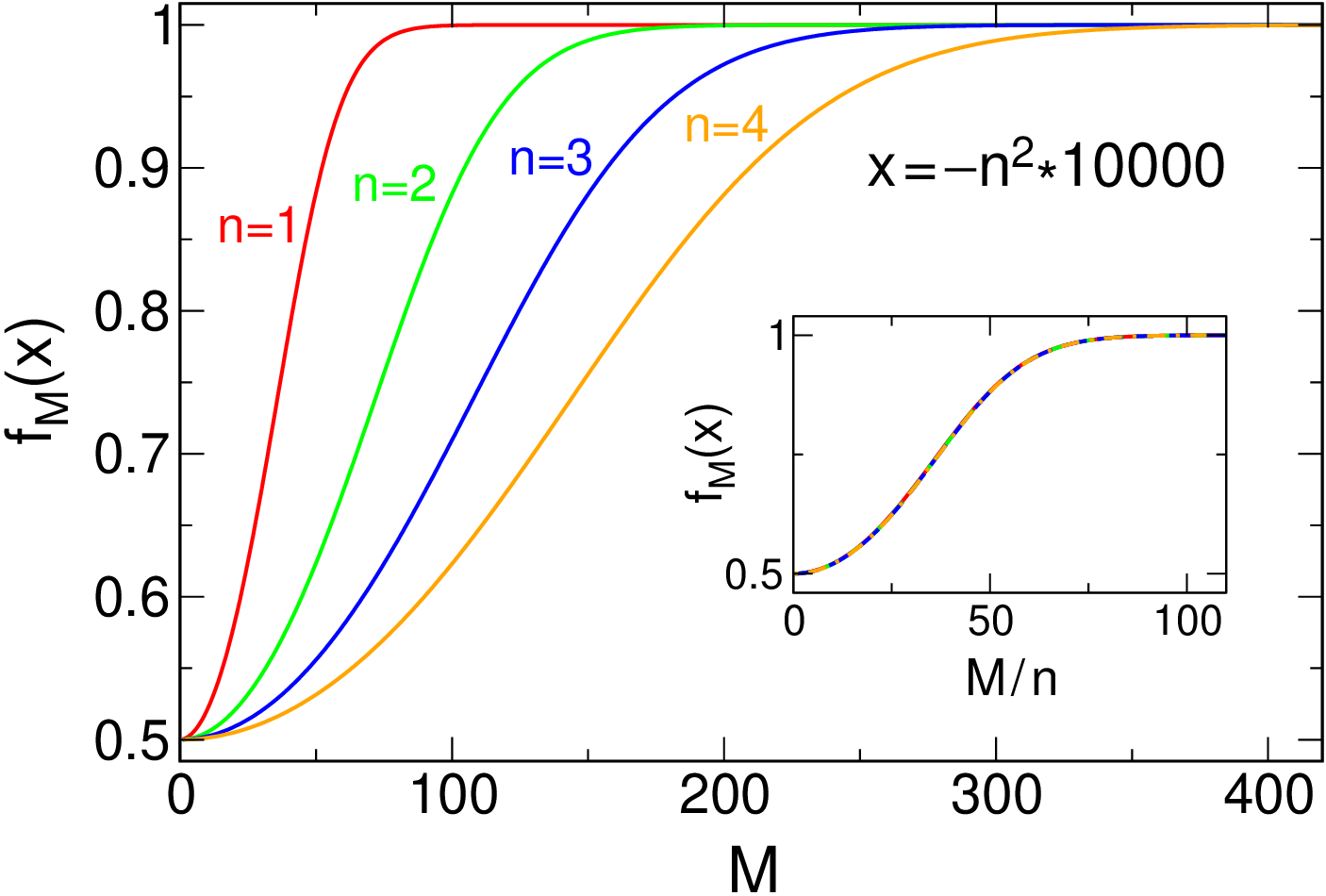}
\caption {Convergence of the continued fraction expansion of the Fermi function to $f_M=1$ for very large negative arguments $x$ as a function of the `chain length' $N=2M$. \textit{Inset:} By rescaling with $1/n$, the four curves collapse, indicating that the truncation $M$ has to increase only as $\sqrt{|x|}$ in order to obtain the same convergence.}
\label{fig:scaling}
\end{figure}

\subsection{Convergence}
Ozaki pointed out the enormous acceleration of convergence one obtains for not too large arguments $|x|$ by replacing the infinite chain by one of finite length $N$ which corresponds to a \textit{finite termination of the infinite continued fraction} (see also Ref.~\onlinecite{hm}). In order to keep the property that all eigenvalues of $B$ come in pairs $(b_\alpha, -b_\alpha)$, one always truncates at an even number $N=2M$ of sites ($M\in\mathbb{N}$).

The eigenvalue problem posed by Eqs.~(\ref{eq:b}) and (\ref{eq:ew}) can be solved numerically using standard routines for all values $M$ of interest. It turns out\cite{Ozaki,hmfn} that the lowest $60$ percent of the $b_\alpha$ are very close to the Matsubara values $[\pi(2n-1)]^{-1}$, and the associated residues are approximately one. The other inverse eigenvalues increase much faster than $\pi(2n-1)$, and their residues $R_\alpha$ become large. As $b_\alpha\in\mathbb{R}$, all poles of the function $f_M(x)$ obtained from restricting the sum in Eq.~(\ref{eq:result2}) to $M$ terms are located on the imaginary axis. This is an important property for the applications discussed in Section \ref{sec:siam} which is not shared by another recent partial fraction decomposition of the Fermi function.\cite{Saalmann}

Already for rather small values of $M$, $f_M(x)$ yields an excellent approximation for $f(x)$ for not too large values of $|x|$. This is illustrated in Fig.~\ref{fig:fermi}. Even at $M=3$, the approximation $f_3(x)$ to the Fermi distribution is excellent up to $|x|\lesssim10$. Likewise, $f_6(x)$ coincides with $f(x)$ to the drawing accuracy for arguments $|x|\lesssim60$. For very large values of $|x|$, all functions $f_M$ eventually tend to $f_M=1/2$.

In Fig.~\ref{fig:scaling} we show the convergence of $f_M$ to one for very large negative values of $x$. Results are shown for four different arguments, increasing in absolute value proportional to $n^2$. By rescaling with $1/n$, the four curves collapse, indicating that $M$ has to increase only as $\sqrt{|x|}$ in order to obtain the same rate of convergence. In passing, we note that it might be possible to analytically analyze the truncation error along the lines of Ref.~\onlinecite{hm}.

\section{Stability of interpolation and continuation of numerical data}
\label{sec:pade}
The linear-response finite-temperature conductance of the single impurity Anderson model\cite{siam} is determined by an energy integral of the derivative of $f(\beta\omega) = 1/[\exp(\beta\omega)+1]$ multiplied by the local single-particle spectral function $\rho(\omega)$:\cite{wingreen}
\begin{equation}\label{eq:lw0}
G = -\Gamma \int \rho(\omega)\,\partial_\omega f(\beta\omega) d\omega =
2\Gamma T \sum_{\alpha>0} R_\alpha\, \tn{Im}\, \frac{d\mc G(i\tilde\omega_\alpha)}{d\tilde\omega_\alpha}~,
\end{equation}
where we have chose units of $\hbar=e=1$ (with $e$ being the elementary charge) and defined the modified Matsubara frequencies $\tilde\omega_\alpha=T/b_\alpha$. The prefactor $\Gamma$ is a measure for the level-lead couplings (see below). The second equality of Eq.~(\ref{eq:lw0}) follows from rotating to the imaginary axis using Eq.~(\ref{eq:result2}) -- it will be derived explicitly in Section \ref{sec:siam}. If $\mc G(i\omega_n)$ is only known numerically at the physical Matsubara frequencies $\omega_n$, both $\rho(\omega)$ and $\mc G'(i\tilde\omega_\alpha)$ need to be determined, e.g., from computing the Pad\'e approximation\cite{pade1,pade2} to $\mc G(i\omega_n)$. Whereas it seems reasonable to expect interpolation (required for extracting $\mc G'(i\tilde\omega_\alpha)$) to be rather stable, an analytic continuation (necessary to calculate $\rho(\omega)$) is in general an ill-controlled procedure (see, e.g., Ref.~\onlinecite{frequenzen}). In this Section, we address this issue explicitly by considering an infinite tight-binding chain with a single site impurity as an exactly-solvable test system. The latter constitutes a special realization of the single impurity Anderson model in absence of Coulomb interactions and is associated with a rich local density of states (featuring both a continuum as well as one bound state).

\begin{figure}[t]
\centering
\includegraphics[width=0.9\linewidth,clip]{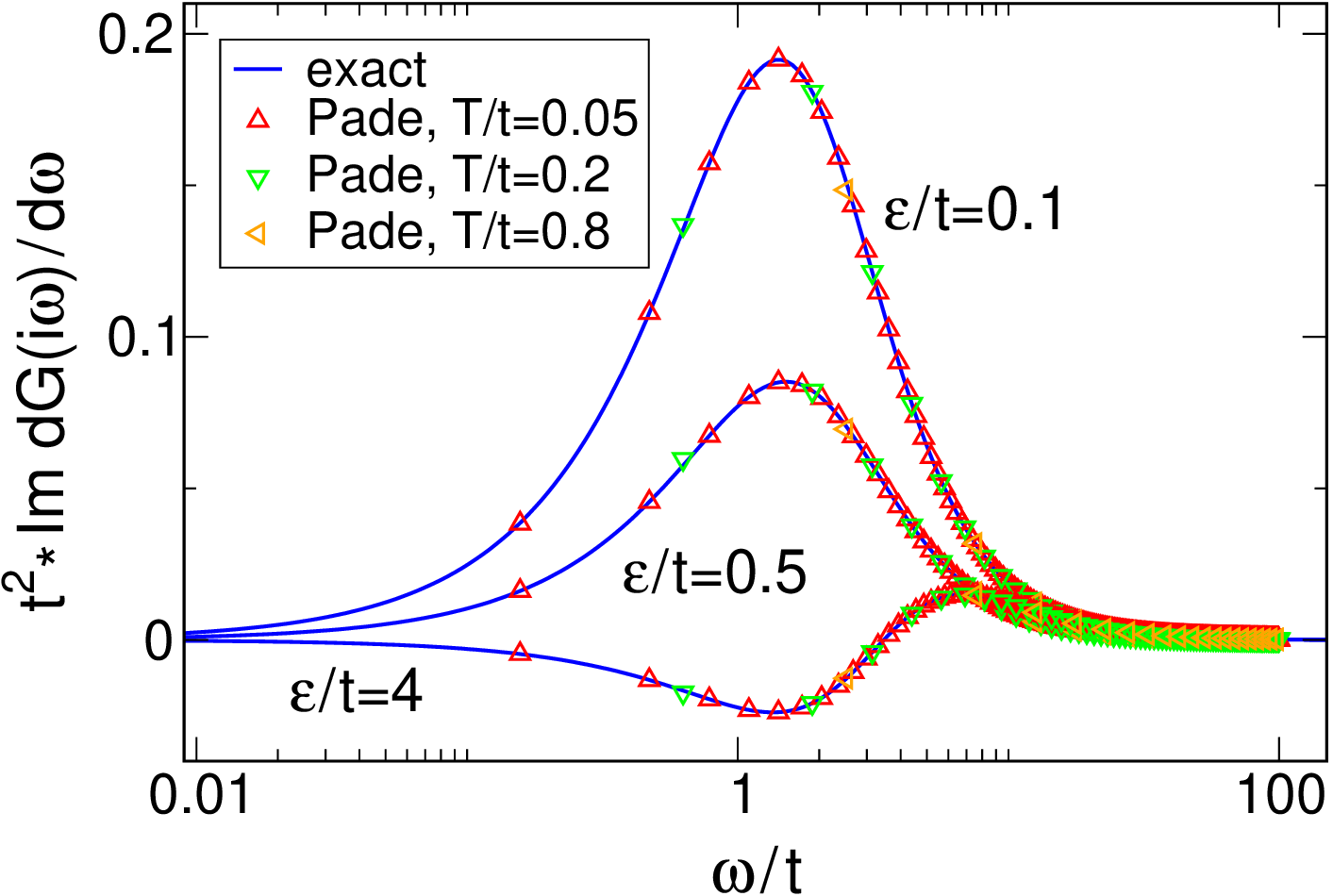}
\caption {(Color online) Derivative of the Matsubara Green function associated with an infinite tight-binding chain with nearest neighbor hopping $t$ and a single site impurity of strength $\epsilon$. Solid lines show the exact result of Eq.~(\ref{eq:gfderiv}), symbols data obtained from computing a Pad\'e approximation to $\mc G(i\omega_n)$ at different temperatures (and thus different physical Matsubara frequencies $\omega_n=(2n+1)\pi T$).\cite{commentdiscr,commentdiscr2} The curves at $\epsilon/t=0.1$ and $\epsilon/t=4$ were both multiplied by a factor of two for clarity. }
\label{fig:int}
\end{figure}

The infinite tight-binding geometry is governed by the Hamiltonian
\begin{equation}
H = -t \sum_i \left(c_{i+1}^\dagger c_i^{\phantom{\dagger}}+\tn{H.c.}\right) + \epsilon c_0^\dagger c_0^{\phantom{\dagger}}~,
\end{equation}
where $t>0$ is the hopping amplitude between nearest neighbors, and $\epsilon$ denotes the strength of an impurity located at site $i=0$. We assume the system to be in thermal equilibrium associated with temperature $T$ and zero chemical potential. Using standard projection techniques,\cite{taylor} the local impurity Green function can be calculated as
\begin{equation}\label{eq:gmat}
\mc G(i\omega) = \frac{1}{i\omega - \epsilon - 2t^2g_0(i\omega)}
\stackrel{\omega>0}{=} \frac{1}{i\sqrt{4t^2+\omega^2}-\epsilon} ~,
\end{equation}
where $g_0(i\omega)$ is the local propagator of an isolated semi-infinite chain:
\begin{equation}
g_0(i\omega) = \frac{1}{2t^2}\left[i\omega-i\tn{sgn}(\omega)\sqrt{4t^2+\omega^2}\right]~.
\end{equation}
The imaginary part of the derivative of $\mc G$ (which determines the linear conductance; see Eq.~(\ref{eq:lw0})) reads
\begin{equation}\label{eq:gfderiv}
\tn{Im}\,\frac{d\mc G(i\omega)}{d\omega} = \frac{\omega\left(4t^2+\omega^2-\epsilon^2\right)}{\sqrt{4t^2+\omega^2}\left(4t^2+\omega^2+\epsilon^2\right)^2}~,
\end{equation}
and the retarded Green function can be calculated straightforwardly from Eq.~(\ref{eq:gmat}) by replacing $i\omega\to\omega+i\eta$:
\begin{equation}
G^{\tn{ret}}(\omega) = \begin{cases}
\frac{1}{i\sqrt{4t^2-\omega^2}-\epsilon} & |\omega|\leq 2t \\
\frac{1}{\tn{sgn}(\omega)\sqrt{\omega^2-4t^2}-\epsilon+i\eta} & |\omega|> 2t~.
\end{cases}
\end{equation}
Thus, the local density of states $\rho(\omega)=-\tn{Im}\,G^\tn{ret}(\omega)/\pi$ is given by
\begin{equation}\label{eq:rho}\begin{split}
\rho(\omega)
= &\frac{1}{\pi}\frac{\sqrt{4t^2-\omega^2}}{4t^2-\omega^2+\epsilon^2}\Theta(|\omega|-2t)\\
& ~~~~+ \frac{\epsilon}{\sqrt{4t^2+\epsilon^2}}\,\delta\big(\omega - \tn{sgn}(\epsilon)\sqrt{4t^2+\epsilon^2}\big)~.
\end{split}\end{equation}
It features a continuum as well as a single pole.

\begin{figure}[t]
\centering
\includegraphics[width=0.9\linewidth,clip]{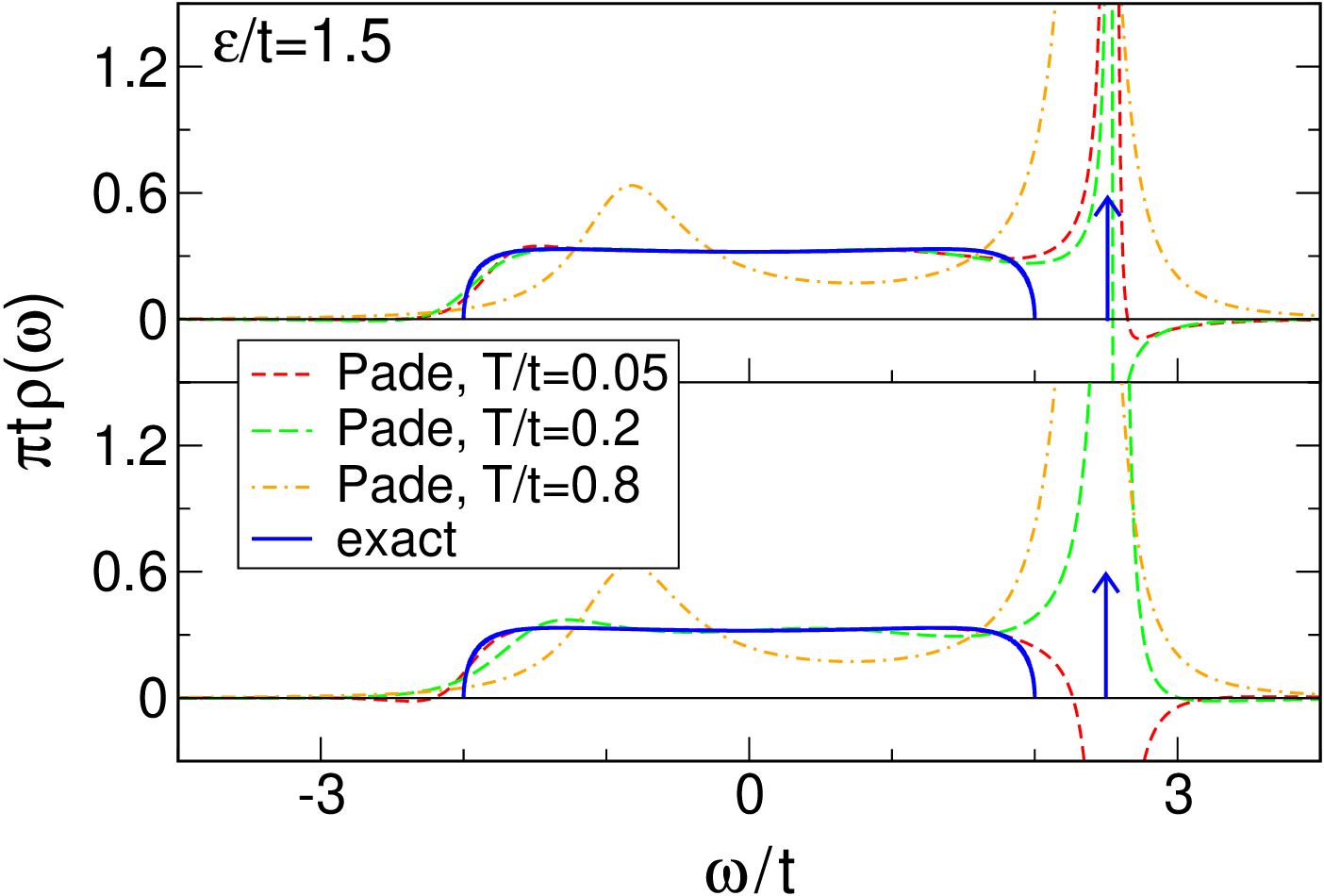}
\caption {(Color online) Tight-binding chain, but now comparing the exact spectral function of Eq.~(\ref{eq:rho}) with the one obtained from Pad\'e approximation. The position and spectral weight of the bound state is indicated by an arrow. Upper and lower panel show results computed with different discretizations of the imaginary frequency axis.\cite{commentdiscr} }
\label{fig:pade}
\end{figure}

We can now test the stability of both the interpolation on the imaginary axis and continuation to real frequencies of the Matsubara Green function. To this end, we compute the Pad\'e approximation (as outlined in Refs.~\onlinecite{pade1,pade2,frequenzen}) to $\mc G(i\omega_n)$ and from that both $\mc G'(i\omega_n)$ as well as $\rho(\omega)$ and compare with the exact results of Eqs.~(\ref{eq:gfderiv}) and (\ref{eq:rho}), respectively. The Pad\'e approximation is calculated numerically for different meshes $\omega_n=(2n+1)\pi T$ of Matsubara frequencies determined by the temperature $T$ (as well as additional discretization parameters\cite{commentdiscr}). For our non-interacting tight-binding chain, both the (exact) local density of states and the derivative $\mc G'(i\omega_n)$ do not depend on $T$.

The results for $\mc G'(i\omega_n)$ and $\rho(\omega)$ obtained from the Pad\'e approximation to the propagator of Eq.~(\ref{eq:gmat}) at different temperatures are shown in Figs.~\ref{fig:int} and \ref{fig:pade}, respectively. It turns out that the former agrees well with the exact curve and is in particular independent of $T$. In contrast, the Pad\'e data for $\rho(\omega)$ depends strongly on temperature (as well as on other discretization parameters\cite{commentdiscr}) and is plagued by severe artifacts such as $\rho(\omega)<0$. As expected, analytic continuation of a numerically known Matsubara Green function is much more unstable than interpolating on the imaginary axis. This demonstrates the advantage of evaluating physical quantities (e.g., the linear conductance which is determined either by $\mc G'(i\tilde\omega_\alpha)$ or $\rho(\omega)$) directly in Matsubara frequency space.

\section{Functional RG for the single impurity Anderson model}
\label{sec:siam}
In this Section, we apply the formalism introduced above to compute the linear conductance $G$ of the single impurity Anderson model (SIAM) at finite temperatures. To this end, we show how to express $G$ in terms of the Matsubara Green function using the continued fraction expansion of the Fermi function and demonstrate that in the exactly-solveable non-interacting (test) case, only a few poles ($M\approx10$) need to be accounted for in order to obtain the conductance in agreement with the one computed on the real axis. For finite (and not too large) Coulomb repulsions, the functional renormalization group -- which in equilibrium is most easily implemented in Matsubara frequency space -- allows for calculating $G$ in good agreement with NRG data. This was previously not possible due to the instability of the analytic continuation.\cite{commentfrg}

\subsection{Model Hamiltonian and linear conductance from the imaginary axis}
The Anderson model is governed by the Hamiltonian\cite{siam}
\begin{equation}\begin{split}
H = & ~\epsilon  \sum_\sigma d_\sigma^\dagger d_\sigma^{\phantom{\dagger}}
+ U\left(d_\uparrow^\dagger d_\uparrow^{\phantom{\dagger}} - \frac{1}{2}\right)
\left(d_\downarrow^\dagger d_\downarrow^{\phantom{\dagger}} - \frac{1}{2}\right) \\
& + \sum_{k\sigma}\epsilon_k\left(c_{Lk\sigma}^\dagger c_{Lk\sigma}^{\phantom{\dagger}}+
c_{Rk\sigma}^\dagger c_{Rk\sigma}^{\phantom{\dagger}} \right) \\
& + \frac{t}{\sqrt{N}}  \sum_{k\sigma} \left(c_{Lk\sigma}^\dagger d_\sigma^{\phantom{\dagger}}+c_{Rk\sigma}^\dagger d_\sigma^{\phantom{\dagger}}+\tn{H.c.}\right) ~,
\end{split}\end{equation}
where $c_{sk\sigma}$ and $d_\sigma$ denote fermionic annihilation operators associated with (left and right) baths as well as a single impurity, respectively. The latter features an onsite energy $\epsilon$ and a Coulomb repulsion $U$ between electrons of different spin directions $\sigma$. The baths are assumed to be in the wide-band limit leading to an energy-independent hybridization $\Gamma=2\pi\rho_\tn{bath} t^2$.

Given the exact local single-particle spectral function $\rho(\omega)$, the linear-response conductance of the SIAM at temperature $T$ can be calculated as\cite{wingreen} (in units of $\hbar=e=1$)
\begin{equation}\label{eq:lw1}
G = -\Gamma \int \rho(\omega)\,\partial_\omega f(\beta\omega) d\omega~,
\end{equation}
where $f(\beta\omega)=1/[\exp(\beta\omega)+1]$ denotes the Fermi distribution. Employing the expansion of Eq.~(\ref{eq:result2}), one can recast Eq.~(\ref{eq:lw1}) in terms of the derivative of the Matsubara Green function:
\begin{equation}\label{eq:lw2}\begin{split}
G & = \Gamma\int f(\beta\omega)\,\partial_\omega\rho(\omega)\,d\omega \\
& =\Gamma \int\left[\frac{1}{2}-T\sum_{\alpha>0}
\left(\frac{R_\alpha}{\omega-i\tilde\omega_\alpha} + \tn{c.c.}\right)\right] \,\partial_\omega\rho(\omega)\,d\omega \\
& =\Gamma T \sum_{\alpha>0} \frac{R_\alpha}{2\pi i}
\int \left(\frac{1}{\omega-i\tilde\omega_\alpha} + \tn{c.c.}\right)\\
& \hspace*{3cm}\times \partial_\omega\left[G^\tn{ret}(\omega)-G^\tn{adv}(\omega)\right]\,d\omega \\[1ex]
& = 2\Gamma T \sum_{\alpha>0} R_\alpha\, \tn{Im}\, \frac{d\mc G(i\tilde\omega_\alpha)}{d\tilde\omega_\alpha}~,
\end{split}\end{equation}
where we have exploited that the retarded and advanced Green functions $G^\tn{ret/adv}$ are analytic in the upper and lower half of the complex plane, respectively.

\begin{figure}[t]
\centering
\includegraphics[width=0.9\linewidth,clip]{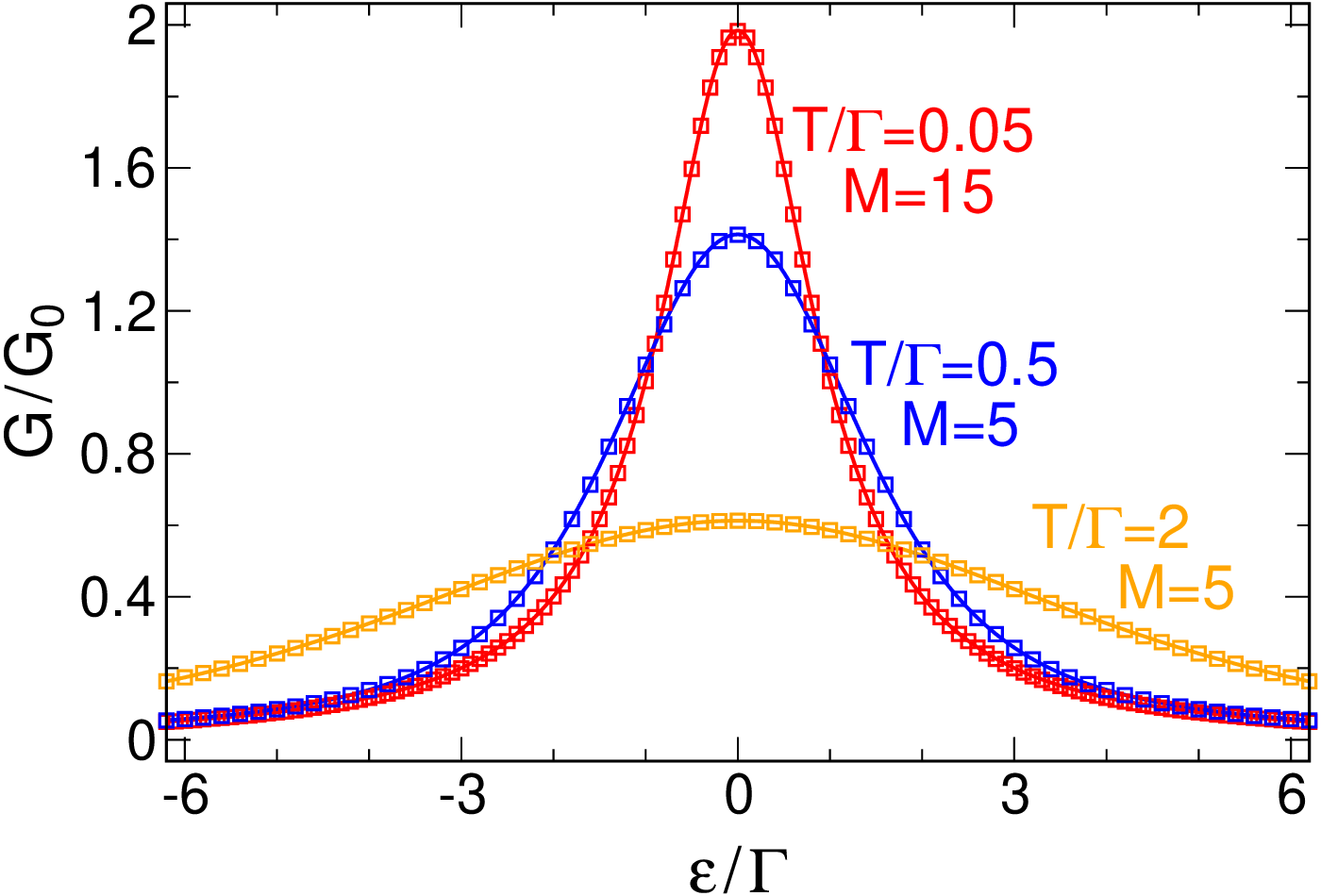}
\caption {(Color online) Linear-response conductance (normalized to the unitary value $G_0=1/(2\pi)$) of the non-interacting single impurity Anderson model as a function of the level position $\epsilon$ and at different temperatures $T$. Solid lines show the exact result calculated from the real axis (Eq.~(\ref{eq:lw1})), symbols were obtained from the imaginary axis using the continued fraction representation of the Fermi function (Eq.~(\ref{eq:lw2})). The expansion was truncated with rather small values of $M$ but agrees nicely with the exact data nevertheless.}
\label{fig:lw1}
\end{figure}

\begin{figure*}[t]
\includegraphics[height=5.1cm,clip]{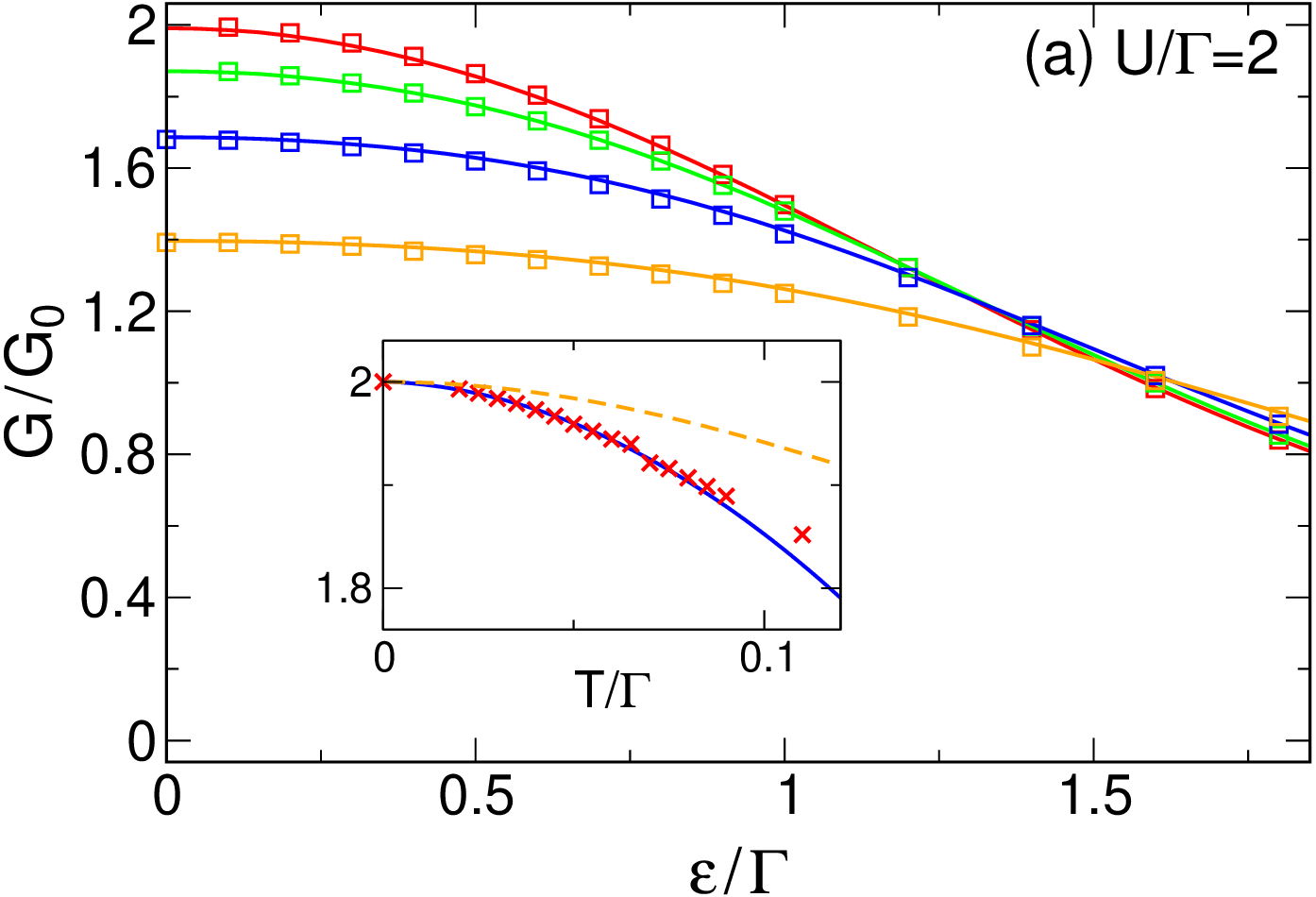}\hspace*{1cm}
\includegraphics[height=5.1cm,clip]{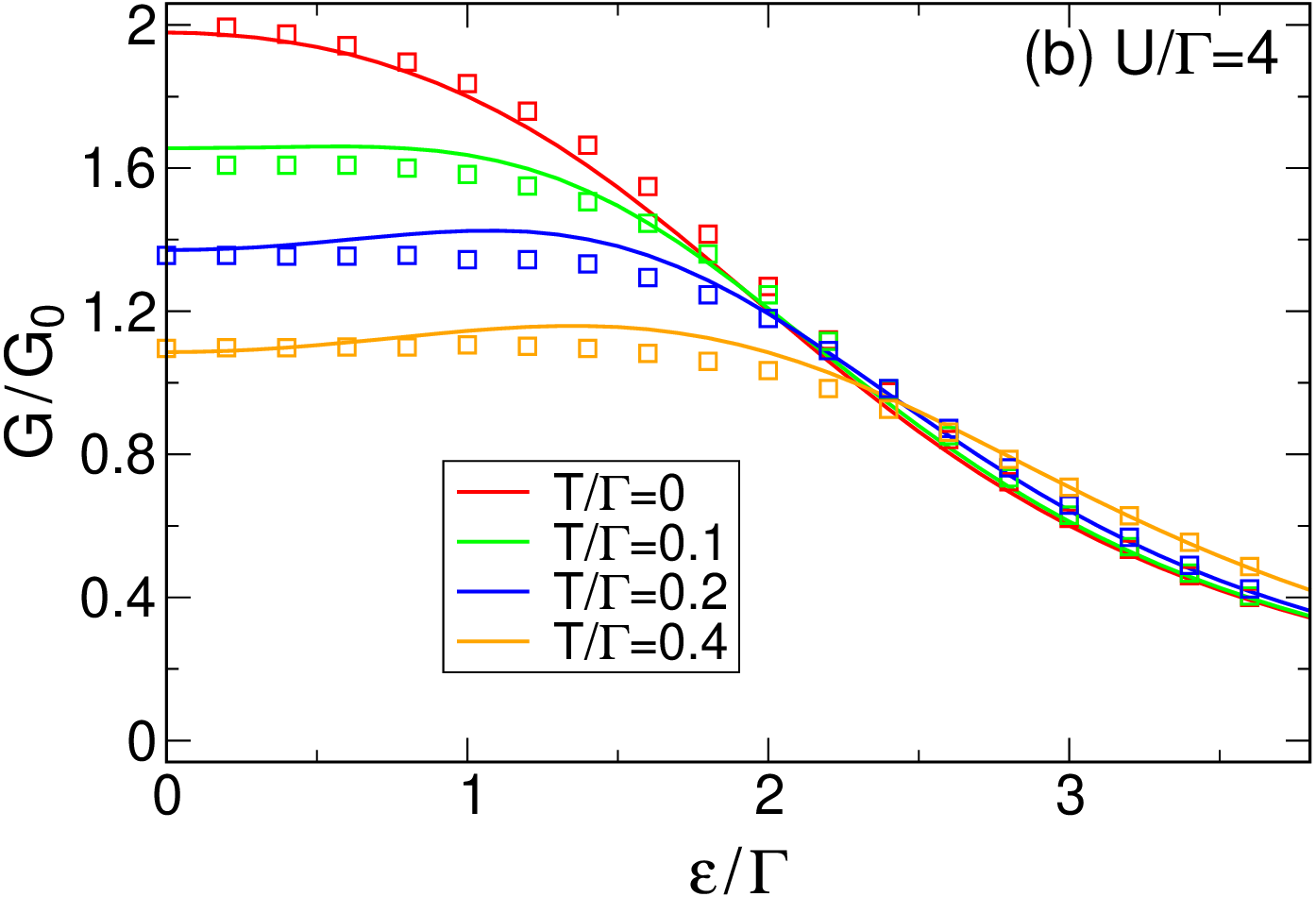}
\caption{(Color online) Linear conductance of the single impurity Anderson model at finite Coulomb interactions and different temperatures $T=0$, $T/\Gamma=0.1$, $T/\Gamma=0.2$, and $T/\Gamma=0.4$ (from top to bottom at $\epsilon=0$). Solid lines show `numerically exact' data obtained from the NRG framework,\cite{thomasnrg} symbols functional RG results computed using the Matsubara FRG scheme outlined in Ref.~\onlinecite{frequenzen} as well as the continued fraction expansion of the Fermi function which underlies Eq.~(\ref{eq:lw2}). \textit{Inset to (a):} Linear conductance as a function of temperature at particle-hole symmetry. Symbols denote FRG data at $U/\Gamma=2$, the solid line a quadratic fit to the latter. The dashed line displays the non-interacting curve. }
\label{fig:lw2}
\end{figure*}

\subsection{Results}
For the \textit{non-interacting case}, both $\rho(\omega)$ and the Matsubara Green function can be computed exactly using standard projection\cite{taylor} or equation of motion\cite{pinkbook} techniques. They read
\begin{equation}\begin{split}
\mc G(i\omega) & = \frac{1}{i\omega-\epsilon+ i\,\tn{sgn}(\omega)\Gamma}~,\\
\rho(\omega) & = \frac{1}{\pi}\frac{\Gamma}{(\omega-\epsilon)^2 + \Gamma^2}~,
\end{split}\end{equation}
and the linear-response conductance $G$ as a function of the level position $\epsilon$ and temperature $T$ can be calculated (numerically) from Eqs.~(\ref{eq:lw1}) and (\ref{eq:lw2}), respectively. Comparing both approaches provides another demonstration of the rapid convergence of the imaginary-axis expression based on the continued fraction expansion of the Fermi function. It turns out (see Fig.~\ref{fig:lw1}) that if the series of Eq.~(\ref{eq:lw2}) is truncated with (rather small) values $M\approx10$, the resulting conductance $G(\epsilon)$ nicely agrees with that of Eq.~(\ref{eq:lw1}) for arbitrary temperatures from $T\ll\Gamma$ to $T\gtrsim\Gamma$.

In \textit{presence of Coulomb interactions}, `numerically exact' results for the spectral function of the SIAM can be obtained from the NRG framework.\cite{nrg} If one aims at describing more complex geometries, however, it is desirable to devise approximate schemes. A recent approach is provided by the Matsubara functional renormalization group\cite{salmhofer} which re-formulates a given many-particle problem in terms of an infinite set of coupled flow equations for single-particle irreducible vertex functions with an infrared cutoff as the flow parameter. Truncation of this hierarchy renders the FRG approximate w.r.t.~the two-particle interaction and can hence \textit{a priori} be justified only in the limit of small $U$. Despite this fact, application of the most simple truncation scheme -- which yields flow equations for effective system parameters and can thus be regarded as a kind of `RG enhanced Hartree-Fock' approach -- to various quantum dot geometries in equilibrium turned out to give accurate results for the \textit{zero-temperature linear conductance} even at fairly large Coulomb interactions.\cite{dotsystems} If one employs a more elaborate truncation (which accounts for the frequency dependence of the two-particle vertex), one can in principle compute the spectral function of the SIAM in agreement with NRG data for small to intermediate values of $U$.\cite{frequenzen,ralf} However, the latter requires an analytic continuation of the Matsubara Green function,\cite{commentfrg} which for the SIAM was observed to be particularly ill-controlled if both $T\neq0$ and $\epsilon\neq0$.\cite{frequenzen} For this reason, it was previously not possible to address the \textit{linear conductance as a function of the level position at finite temperatures}. The continued fraction expansion for the Fermi function now allows for calculating $G(\epsilon,T)$ directly from the imaginary axis. To this end, we extract the FRG approximation\cite{frgappr3} to the Matsubara Green function of the SIAM using precisely the formalism outlined in Ref.~\onlinecite{frequenzen}. Thereafter, we compute the derivative $\mc G'(i\tilde\omega_\alpha)$ from interpolating by virtue of a Pad\'e approximation -- which again turns out to be far more stable than continuation to the real axis -- and ultimately the linear conductance by Eq.~(\ref{eq:lw2}). The result is shown (and compared to NRG data\cite{nrg}) in Fig.~\ref{fig:lw2}. The FRG correctly describes the widening of the Lorentzian lineshape of $G(\epsilon)$ due to Coulomb correlations (signaling the development of a Kondo plateau) and for small $U/\Gamma=2$ (intermediate $U/\Gamma=4$) agrees nicely (decently) with the NRG reference. Moreover, the temperature-dependence of $G(T)$ at particle-hole symmetry $\epsilon=0$ is purely quadratic (see the inset to Fig.~\ref{fig:lw2}(a)) as expected from Fermi-liquid theory.\cite{hewson} For larger values of $U$, the agreement with the numerical RG data becomes worse (which is in line with the results of Ref.~\onlinecite{frequenzen}).

\section{Conclusion}
In this Paper we illustrated how to compute the finite-temperature linear-response conductance $G$ of quantum impurity models from the Matsubara Green function $\mc G(i\omega_n)$ using a rapidly-converging continued fraction expansion of the Fermi function recently introduced by Ozaki. In case that $\mc G(i\omega_n)$ is only known numerically, this formalism allows to circumvent the need to carry out an (ill-controlled) analytic continuation to the real axis. As an application, we studied the single impurity Anderson model at finite temperatures within the framework of the Matsubara functional renormalization group and showed that $G$ can be obtained accurately in comparison with numerical RG data for not too large Coulomb interactions. The FRG therefore holds the promise for future treatments of more complex quantum dot geometries which cannot be easily addressed within the NRG framework.

\section*{Acknowledgments}
We are grateful to the Deutsche Forschungsgemeinschaft for support via FOR723 and thank Theo Costi for providing his code to carry out NRG calculations. Useful discussions with Jan von Delft and Harmut Monien are acknowledged.


\begin{thebibliography}{99}
\bibitem{commentdiscr}
For finite temperatures, the Matsubara frequencies are discrete but extend to infinity nevertheless. Thus, one needs to introduce a finite frequency mesh in order to numerically compute the Pad\'e approximation to the Matsubara Green function. There are various possibilities for realizing such mesh (e.g., linearly or `logarithmically' spaced groups of frequencies). While the derivative $\mc G'(i\omega_n)$ turns out to be independent of those details, the local density of states is rather `unstable'. The latter is illustrated in Fig.~\ref{fig:pade} where the mesh employed is the one detailed in Ref.~\onlinecite{frequenzen}. Following the notation of this Ref.~\onlinecite{frequenzen}, the discretization parameters $\{N,N_0,S,A\}$ of the upper panel are given by $T/\Gamma=0.05: \{41,11,2,2\}$, $T/\Gamma=0.2: \{41,12,2,2\}$, and $T/\Gamma=0.8: \{41,20,2,2\}$, and those of the lower panel read $T/\Gamma=0.05: \{61,15,2,2\}$, $T/\Gamma=0.2: \{61,22,2,2\}$, and $T/\Gamma=0.8: \{61,61,2,2\}$.
\bibitem{commentdiscr2}
The symbols in Fig.~\ref{fig:int} show the Pad\'e approximation for the derivative $\mc G'(i\omega_n)$ only at the physical Matsubara frequencies $\omega_n$. If one seeks to calculate the linear-response conductance by virtue of the continued fraction expansion of the Fermi function (e.g., using Eq.~(\ref{eq:lw0}) for the SIAM), it is necessary to know $\mc G'(i\tilde\omega_\alpha)$ at the modified frequencies $\tilde\omega_\alpha$. As discussed in Ref.~\onlinecite{Ozaki}, the first 60 percent of the $\tilde\omega_\alpha$ almost coincide with their Matsubara counterparts, and it turns out that $\mc G'(i\omega)$ can be computed in a stable way for arbitrary arguments $\omega$ except those close to the real axis ($\omega<\omega_0$).
\bibitem{commentfrg}
If one implements the FRG in Keldysh space, the finite-temperature linear-response (as well as the non-equilibrium) conductance of the SIAM can be obtained without carrying out an analytic continuation.\cite{severinsiam} The structure of the FRG flow equations, however, is far more complex in the Keldysh than in the Matsubara formalism. If one is interested in linear response only, the latter thus provides the most simple framework to study more complex geometries.
\bibitem{nrg}
Standard NRG is a well-established numerical tool to compute low-energy equilibrium properties of quantum impurity systems. A detailed introduction to this method can be found in Ref.~\onlinecite{thomasnrg}.
\bibitem{frgappr3}
More precisely, we employ the so-called `approximation 1' in the notation of Ref.~\onlinecite{frequenzen}.
\end{thebibliography}
\end{document}